\documentclass[a4paper,11pt]{article}
\pdfoutput=1 % if your are submitting a pdflatex (i.e. if you have
\usepackage{jcappub} % for details on the use of the package, please
\usepackage[english]{babel}
\usepackage[utf8]{inputenc}
\usepackage[T1]{fontenc}

\usepackage{xcolor}
\usepackage{float}
\usepackage{cancel}
\usepackage{dblfloatfix}
\usepackage[symbol]{footmisc}
\usepackage{graphicx}

\usepackage{amsmath}
\usepackage{amssymb}
\usepackage{slashed}
\usepackage{braket}

\newcommand{\dd}{\mathrm{d}}
\newcommand{\defEq}{\overset{\mathrm{def}}{=}}
\newcommand{\MeV}{\ \mathrm{MeV}}
\newcommand{\GeV}{\ \mathrm{GeV}}
\newcommand{\VEV}{\mathrm{vev}}
\newcommand{\Vct}{V_{\mathrm{CT}}}
\newcommand{\Vcw}{V_{\mathrm{CW}}}

\newcommand{\distTwo}{\hspace*{2mm} , \hspace*{2mm}}
\newcommand{\dist}{\hspace*{4mm} , \hspace*{4mm}}

\title{Dark Matter and Nature of Electroweak Phase Transition with an Inert Doublet}

\author[a]{Sven Fabian,}
\author[a]{Florian Goertz,}
\author[b,a]{and Yun Jiang}

\affiliation[a]{Max-Planck-Institut f{\"u}r Kernphysik, Saupfercheckweg 1, 69117 Heidelberg, Germany}
\affiliation[b]{MOE Key Laboratory of TianQin Mission, TianQin Research Center for Gravitational Physics \& School of Physics and Astronomy, Frontiers Science Center for TianQin, CNSA Research Center for Gravitational Waves, Sun Yat-sen University (Zhuhai Campus), Zhuhai 519082, P. R. China}

\emailAdd{fabian@mpi-hd.mpg.de}
\emailAdd{florian.goertz@mpi-hd.mpg.de}
\emailAdd{jiangyun5@sysu.edu.cn}

\abstract{We provide a comprehensive and up-to-date analysis of the prospects to realize Dark Matter (DM) in the Inert Doublet Model, while simultaneously enhancing the Electroweak Phase Transition (EWPhT) such as to allow for electroweak baryogenesis. Instead of focusing on certain aspects or mass hierarchies, we perform extensive, yet fine-grained, parameter space scans, where we analyze the nature of the EWPhT in both the light and the heavy DM regions, confronting it with the amount of DM potentially residing in the lightest inert-doublet state. Thereby, we point out a viable region where a non-trivial two-step \mbox{EWPhT} can appear, without being in conflict with direct-detection bounds, which could leave interesting imprints in gravitational wave signatures. We propose new benchmarks with this feature as well as update benchmarks with a strong first-order transition in the light of new XENON1T limits. Moreover, taking into account these latest bounds as well as relevant collider constraints, we envisage a region for light DM with a small mass splitting, lifting the usual assumption of exact degeneracy of the new non-DM scalars, such as to avoid collider bounds while providing a fair DM abundance over a rather large DM mass range. This follows from a detailed survey of the impact of co-annihilations on the abundance, dissecting the various channels.}

\begin{document}
\maketitle
\flushbottom

\section{Introduction and Setup}

The presence of a baryon-asymmetric universe and of non-luminous Dark Matter (DM) leads to the conclusion that our current understanding of nature, and thus the Standard Model of Particle Physics (SM), is incomplete. In consequence, although in the past decades the SM was very successful in describing the microscopic properties and interactions of the elementary particles found so far, with the latest highlight being the discovery of the predicted Higgs boson at the LHC in 2012, it needs to be extended to agree with these findings.
On the observational side, the energy budget of the universe can be characterized by the $\Lambda\mathrm{CDM}$ model as the standard picture of Big Bang Cosmology, based on general relativity, which includes the cosmological constant $\Lambda$ and Cold Dark Matter (CDM), and delivers a good fit to the plethora of cosmological observations, starting from six parameters~\cite{PDGLCDM}. However, what is the {\it nature} of DM (or of the cosmological constant) is still unsettled and cannot be addressed within the SM. It is thus hoped that a microscopic theory extending the SM will eventually be unveiled, that could resolve this issues. The same is true for the dynamics behind the observed non-vanishing baryon-density, which cannot be generated via SM physics - lacking a strong first-order Electroweak Phase Transition (EWPhT), to allow for out-of-equilibrium dynamics, and a sufficient amount of CP violation. 

While we will comment more on means to generate additional CP violation later, %with physics beyond the SM (BSM)
the main focus of this article will be to investigate DM Physics as well as the EWPhT dynamics comprehensively in the framework of the Inert Doublet Model (IDM), which furnishes a very clear and predictive extension of the SM. Here, the Higgs sector of the SM is augmented with an electroweak (EW) doublet, odd under a $\mathbb{Z}_{2}$ symmetry and with vanishing vacuum expectation value (vev) at zero temperature, which naturally includes a DM candidate as well as the potential to enhance the EWPhT.
We aim for accommodating the correct DM abundance and the presence of a strong first-order EWPhT simultaneously, thereby going through the various possibilities, while meeting all relevant constraints from collider searches and cosmology. Even though the IDM has been subject to several analyses on DM physics and on different possibilities to realize a strong first-order EWPhT, see Refs.~\cite{IDM,IDMLep2,Dolle:2009fn,Chowdhury:2011ga,Borah:2012pu,Gil:2012ya,Cline:2013bln,Profumo,IDM5}, a comprehensive scan of the whole parameter space regarding the nature of the EWPhT and a successful DM physics, including up-to-date constraints and considering the full range of (DM) masses, would be useful. Moreover, a detailed exploration of the potential to achieve a successful two-step EWPhT and of its particularities is still lacking. Both will be provided in this article, including a detailed treatment of finite temperature effects, employing thermal resummation. 

This article is organized as follows. In the remainder of this section we introduce the model, summarize important constraints from EW precision data and exotic SM Higgs decay as well as analyze the IDM scalar potential at finite temperature. Subsequently, in Sec.~\ref{sec:DM}, we study comprehensively the DM physics over a large range of masses, deriving the relic abundance taking carefully into account the effect of different co-annihilation channels
and incorporating most recent limits from DM direct detection (DD),
which turn out to close recently still available mass ranges. On the other hand, we propose a new viable spectrum with small mass splitting that can lead to a proper amount of DM without being confined to narrow stripes in the DM mass. In Sec.~\ref{sec:EWPhT} we explore various particularities of the EWPhT in the IDM, presenting for the first time a detailed, fine-grained, survey of the parameter space regarding the nature of the EWPhT, which we find to be a two-step transition in well-constrained stripes of parameters, fulfilling all relevant bounds and providing a viable DM abundance. In turn, we define a new set of benchmarks where we trace the EWPhT, arriving at the proper EW vacuum in a non-trivial way, which are expected to lead to interesting gravitational wave (GW) signatures with multi-peaked spectra. Finally, we conclude in Sec.~\ref{sec:concl}.

\subsection{The Inert Doublet Model}

In the IDM~\cite{IDMoriginal1,IDMoriginal2}, an additional EW doublet $H_{2}$ is added to the SM Higgs doublet $H_{1}$, reading
\begin{equation}
\label{eq:doublets}
H_{1} = \frac{1}{\sqrt{2}} \begin{pmatrix}
\sqrt 2 \phi^{+} \\ 
h + i\phi
\end{pmatrix} \dist
H_{2} = \frac{1}{\sqrt{2}}  \begin{pmatrix}
\sqrt{2}H^{+} \\ 
H + iA
\end{pmatrix}\,.
\end{equation}
Here, $h$ is the SM-like neutral Higgs field,\footnote{With some abuse of notation, we will use the same name later for the physical fluctuation around the vev.} with the vev 
$\langle h \rangle \equiv v \approx 246 \GeV$  
at zero temperature, % = \left( \sqrt{2}G_{F} \right)^{-1/2} derived from the Fermi constant $G_{F}$,
$\phi^{+}$, $\phi^{-} = \phi^{+*}$, and $\phi$ are the EW Goldstone bosons, $H^\pm$ and $H$ are new, charged and neutral, CP-even scalars, while $A$ is a new CP-odd scalar (acknowledging that it is actually not possible to unambiguously assign definite CP properties to the neutral states of the second doublet due to the absence of suited processes \cite{IDM}). Imposing a discrete $\mathbb{Z}_{2}$ symmetry, under which $H_2$ is odd while all the SM fields are even, prevents the lightest $\mathbb{Z}_{2}$-odd particle from decaying into SM particles. Thus, it plays the role of the DM candidate in this model. 

The resulting scalar potential reads (see, {\it e.g.}, Ref.~\cite{IDM})
\begin{equation}
\begin{split}
V & =  \mu_{1}^{2} \left\vert H_{1} \right\vert^{2} + \mu_{2}^{2} \left\vert H_{2} \right\vert^{2} + \lambda_{1} \left\vert H_{1} \right\vert^{4} + \lambda_{2} \left\vert H_{2} \right\vert^{4}  \\
& +   \lambda_{3} \left\vert H_{1} \right\vert^{2} \left\vert H_{2} \right\vert^{2} + \lambda_{4} \left\vert H_{1}^{\dagger}H_{2} \right\vert^{2} + \frac{\lambda_{5}}{2} \left[ \left( H_{1}^{\dagger}H_{2} \right)^{2} + \mathrm{h.c.} \right]\,, \label{Eq:Vtree}
\end{split}
\end{equation}
with $\mu_{2}$, $\lambda_{2}$, $\lambda_{3}$, $\lambda_{4}$, $\lambda_{5}$ as free, real, parameters and the remaining Lagrangian coincides with the SM one, up to gauge invariant kinetic terms for the new doublet. In particular,  due to $\mathbb{Z}_2$-invariance, no new Yukawa couplings appear. Using the notations for the Higgs portal couplings
\begin{equation}
\label{eq:lamdef}
\lambda_{345} \defEq \lambda_{3} + \lambda_{4} + \lambda_{5} \dist \bar{\lambda}_{345} \defEq \lambda_{3} + \lambda_{4} - \lambda_{5}=\lambda_{345}-2 \lambda_5
\end{equation}
and employing Landau gauge ($\xi = 0$), the mass matrices for the neutral scalars, $h$ and $H$, the pseudoscalar states, $\phi$ and $A$, and for the charged scalars, $\phi^\pm$ and $H^{\pm}$, are given by 
\begin{equation}
\label{eq:MM}
M_{S}^{2} = \begin{pmatrix}
2\lambda_{1}v^{2} & 0 \\ 
0 & \mu_{2}^{2} + \lambda_{345}v^{2}/2 
\end{pmatrix},\
M_{P}^{2} = \frac{1}{2} \begin{pmatrix}
0 & 0 \\ 
0 & \bar{\lambda}_{345}v^{2} + 2\mu_{2}^{2}
\end{pmatrix},\
M_{\pm}^{2} = \frac{1}{2} \begin{pmatrix}
0 & 0 \\ 
0 & \lambda_{3}v^{2} + 2\mu_{2}^{2}
\end{pmatrix}, 
\end{equation}
from which the mass eigenvalues can be read off trivially\footnote{For calculating the one-loop potential, these matrices will later by generalized with (non-diagonal) field-dependent entries.}. We note that the portal parameters can be expressed in terms of these masses and $\lambda_{345}$ via
\begin{equation}
\label{eq:lambdas}
\lambda_{3} = \lambda_{345} + 2\frac{m_{H^{\pm}}^{2} - m_{H}^{2}}{v^{2}} \dist \lambda_{4} = \frac{m_{A}^{2} + m_{H}^{2} - 2m_{H^{\pm}}^{2}}{v^{2}} \dist \lambda_{5} = \frac{m_{H}^{2} - m_{A}^{2}}{v^{2}} \ ,
\end{equation}
such that, eliminating also $\mu_2$ via Eq.~(\ref{eq:MM}), we are left with the new set of free parameters $\{\lambda_2,\lambda_{345},m_H,m_{H^{\pm}},m_A\}$, which we will employ in the following.

We already observe that, due to the new couplings being real, the IDM alone does not induce additional CP violation, which is however required in order to allow for EW baryogenesis \cite{Sakharov:1967dj}. In consequence, we envisage additional CP violating sources, which we think of agnostically as higher-dimensional operators in an effective field theory (EFT), sticking to the field content introduced in Eq.~(\ref{eq:doublets}).\footnote{For explicit extensions of the IDM with further scalars to allow for CP violation see, \emph{e.g.}, Refs.~\cite{Grzadkowski:2010au,Krawczyk:2015xhl,Cordero-Cid:2020yba}.} Some possible operators are collected in
\begin{equation}
\label{eq:CPo}
\begin{split}
{\cal L}_{\rm \cancel{CP}} \supset\ &   C_{H_{1} \tilde F} \left\vert H_{1}\right\vert^2 \tilde F_{\mu\nu}^I F^{I\,\mu\nu} %+ C_{H \tilde B} |H_1|^2 \tilde B_{\mu\nu} B^{\mu\nu} + C_{H \tilde W B} H_1^\dagger \sigma _I H_1 \tilde W_{\mu\nu}^I B^{\mu\nu}
+ C_{qH_{1}} \left\vert H_{1} \right\vert^2 \bar q_L H_{1}  q_R +  C_{H_2 \tilde F} |H_2|^2 \tilde F_{\mu\nu}^I F^{I\,\mu\nu} 
\,,
\end{split}
\end{equation}
with $F_{\mu\nu}^I$ denoting EW gauge bosons, where the first two types of operators have been studied extensively in the literature \cite{Dine:1990fj,Dine:1991ck,Balazs:2016yvi,Ferreira:2016jea}\footnote{Beyond that, Ref.~\cite{Balazs:2016yvi} provides a systematic survey of new sources of CP violation in the SMEFT in the context of successful EW baryogenesis.}. Since they both contribute significantly to electric dipole moments (see, \emph{e.g.}, Ref.~\cite{Balazs:2016yvi}), it could be interesting to consider the third operator, which in analogy to the analysis performed in Ref.~\cite{Dine:1990fj} could induce appreciable CP violation at finite temperature for $C_{H_2 \tilde F} \sim ({\rm TeV})^{-2}$.
Finally, we note that a comprehensive study of the IDM could also be considered as a first step to a general survey of the possibility to simultaneously realize DM and EW baryogenesis, in extended scalar sectors - either linking the DM candidate via SU(2) quantum numbers to the SM, as in the IDM, or moving the DM into a singlet-like dark sector, potentially coupled to the SM via a singlet mediator,
where one could also employ a 'model-independent' framework such as the extended Dark Matter EFT \cite{ExtendedDMEFT}.

\subsection{Theoretical Constraints and Bounds from Electroweak Precision and Higgs Data}
\label{sec:ThBounds}
In our analysis of the IDM, we apply theoretical constraints, such as vacuum stability and perturbative unitarity, as well as experimental bounds both from EW precision data and from exotic SM Higgs decays. Vacuum stability requires the four relations~(see, \emph{e.g.}, Ref.~\cite{IDM})
\begin{equation}
\lambda_{1}>0 \dist \lambda_{2}>0 \dist \lambda_{3} > -2\sqrt{\lambda_{1}\lambda_{2}} \dist \lambda_{3} + \lambda_{4} - \left\vert \lambda_{5} \right\vert > -2\sqrt{\lambda_{1}\lambda_{2}} \ ,
\end{equation}
and a charge-breaking vacuum is avoided by satisfying
\begin{equation}
\lambda_{4} - \left\vert \lambda_{5} \right\vert < 0 \ .
\end{equation}
The conditions for perturbative unitarity read $\left\vert c_{i} \right\vert < 8\pi$, with~\cite{IDM}
\begin{equation}
\begin{split}
c_{1,2} = \lambda_{3} \pm \lambda_{4} \dist & c_{3,4} = -3\lambda_{1} - 3\lambda_{2} \pm \sqrt{9 \left( \lambda_{1} - \lambda_{2} \right)^{2} + \left( 2\lambda_{3} + \lambda_{4} \right)^{2}} \ , \\
c_{5,6} = \lambda_{3} \pm \lambda_{5} \dist & c_{7,8} = -\lambda_{1} - \lambda_{2} \pm \sqrt{\left( \lambda_{1} - \lambda_{2} \right)^{2} + \lambda_{5}^{2}} \ ,  \\
c_{9,10} = \lambda_{3} + 2\lambda_{4} \pm 3\lambda_{5} \dist & c_{11,12} = -\lambda_{1} - \lambda_{2} \pm \sqrt{\left( \lambda_{1} - \lambda_{2} \right)^{2} + \lambda_{4}^{2}} \ .
\end{split}
\end{equation}

Moreover, since the decay widths of the EW gauge bosons are measured to high accuracy and no hint for new physics was found in LEP data, potential decays of the $W^{\pm}$ or $Z$ into states of the additional doublet $H_{2}$ are excluded in our analysis by constraining the mass spectra to~\cite{IDM}
\begin{equation}
m_{H} + m_{H^{\pm}} > m_{W^{\pm}} \distTwo m_{A} + m_{H^{\pm}} > m_{W^{\pm}} \distTwo
m_{H} + m_{A} > m_{Z} \distTwo 2m_{H^{\pm}} > m_{Z}\ .
\end{equation}
Finally, a reinterpretation of the LEP-II data gives rise to the exclusion of an intersection of mass ranges, which can be evaded by fulfilling one of the following conditions~\cite{IDM,IDM5,IDMLep2,Pierce_2007}
\begin{equation}
\label{eq:LEPsplit}
m_{H} > 80\GeV \ \ \cup \ \ m_{A} > 100\GeV \ \  \cup \ \  m_{A}-m_{H} < 8\GeV\,,
\end{equation}
in addition to the general bound
\begin{equation}
m_{H^{\pm}} > 70\GeV \ ,
\end{equation}
from searches for charged Higgs pair production.

For a general parametrization of corrections to EW precision observables, the three oblique parameters $S$, $T$ and $U$ have been introduced in Ref.~\cite{STUparameters} and are defined to vanish in the absence of new physics \cite{Baak_2014}. Assuming $U$ to vanish, the other two contributions are given by~\cite{IDMoriginal2,IDM}
\begin{equation}
\label{eq:STU}
\begin{split}
S & =  \frac{1}{72\pi \left( x_{2}^{2} - x_{1}^{2} \right)^{3}} \left[ x_{2}^{6}f_{a}\left( x_{2} \right) - x_{1}^{6} f_{a} \left( x_{1} \right) + 9x_{1}^{2}x_{2}^{2} \left( x_{2}^{2}f_{b} \left( x_{2} \right) - x_{1}^{2}f_{b} \left( x_{1} \right) \right) \right]\\
T & =  \frac{1}{32\pi^{2}\alpha v^{2}} \left[ f_{c} \left( m_{H^{\pm}}^{2}, m_{A}^{2} \right) + f_{c} \left( m_{H^{\pm}}^{2}, m_{H}^{2} \right) - f_{c} \left( m_{A}^{2}, m_{H}^{2} \right)  \right] \\
&\simeq   \frac{1}{24\pi^{2}\alpha v^{2}}(m_{H^\pm}-m_H)(m_{H^\pm}-m_A)\,.
\end{split}
\end{equation}
Here, $\alpha\approx 1/127$ denotes the fine-structure constant at the scale of the $Z$ boson mass,
\begin{equation}
\label{eq:fs}
f_{a} \left( x \right) \defEq -5 + 12\ln x \distTwo
f_{b}\left( x \right) \defEq 3-4\ln x \distTwo
f_{c} \left( x,y\right) \defEq \begin{cases} \frac{x+y}{2} - \frac{xy}{x-y}\ln \frac{x}{y} &\mathrm{for} \ x\neq y \\ 0 &\mathrm{for} \ x=y \end{cases}\,,
\end{equation}
where
\begin{equation}
x_{1} \defEq \frac{m_{H}}{m_{H^{\pm}}} \dist x_{2} \defEq \frac{m_{A}}{m_{H^{\pm}}} \ ,
\end{equation}
and the given approximate relation for the $T$ parameter holds for not too large splittings.
Employing the measured SM Higgs mass $m_{h} = 125 \GeV$, the best fit with $U=0$ yields \cite{Baak_2014}
\begin{equation}
S = 0.06 \pm 0.09 \dist T = 0.10 \pm 0.07 \ .
\end{equation}
From Eqs.~(\ref{eq:STU}) and (\ref{eq:fs}) one can for example inspect that a large splitting between $m_{H^{\pm}}$ and $m_{A}$ would induce sizable corrections to the $T$ parameter, violating the above constraints.

The final constraint arises from the absence of exotic Higgs decays, leading to an upper limit on the branching ratio of~\cite{BRatlas,BRcms} 
\begin{equation}
\mathrm{BR} \left( h \rightarrow \mathrm{inv.} \right) \defEq  \frac{\Gamma \left( h \rightarrow \mathrm{inv.} \right)}{\Gamma \left( h \rightarrow \mathrm{inv.} \right) + \Gamma \left( h \rightarrow \mathrm{SM} \right)} < \begin{cases} 0.26 \ \mathrm{from \ ATLAS} \ \\ 0.19 \ \mathrm{from \ CMS} \  \end{cases}
\end{equation}
%\hspace*{4mm} \mathrm{or} \hspace*{4mm} \mathrm{BR}_{\mathrm{CMS}} \left( h \rightarrow \mathrm{inv.} \right) < 
at $95\%$ confidence level with the decay width
\begin{equation}
\Gamma \left( h \rightarrow \mathrm{inv.} \right) =  \frac{\left( \lambda_{345} m_{W} \right)^{2}}{8\pi g_{W}^{2}m_{h}} \sqrt{1-4\left( \frac{m_H}{m_{h}} \right)^{2}}\,,
\end{equation}
where $g_{W}$ is the weak coupling constant, 
and the theoretical SM Higgs width reads~\cite{PDG} 
\begin{equation}
\Gamma \left( h \rightarrow \mathrm{SM} \right) = 4.07 \MeV^{+4.0\%}_{-3.9\%} \ .
\end{equation}
All these constraints will be implemented in our numerical analysis in Sections \ref{sec:DM} and \ref{sec:EWPhT}.

\subsection{Finite Temperature Effects and Electroweak Phase Transition}

To investigate the EWPhT in the subsequent two-field analysis, concepts of finite-temperature QFT will be employed - but first
we will add the zero-temperature one-loop potential to the tree-level potential in Eq.~(\ref{Eq:Vtree}), with the corresponding Coleman-Weinberg (CW) potential in Landau gauge and $\overline{\mathrm{MS}}$-scheme reading
\begin{equation}
\label{eq:CW}
\Vcw \left( h, H \right) = \sum_{i} \frac{n_{i}}{64\pi^{2}} \hat{m}_{i}^{4} \left( h, H \right) \left[ \ln \left( \frac{\hat{m}_{i}^{2}}{Q^{2}} \right) - C_{i} \right]\,.
\end{equation}
Here, $i=W^\pm,Z,t,h,H,\phi,A,\phi^\pm,H^\pm$, where we just kept the most massive top quark,\footnote{We explicitly checked that the contributions from the bottom quark and the tau lepton are very small.} $Q$ is the renormalization scale, $n_{i}$ denote the number of bosonic and fermionic degrees of freedom, and $C_{i}$ are renormalization-scheme dependent constants. The latter two are given by
$n_W\!=\!6,\,n_Z\!=\!3,\,n_t\!=\!-12,\,n_\Phi\!=\!1,\,n_{\Phi^\pm}\!=\!2$ and $C_W\!=\!C_Z\!=\!5/6,\,C_t\!=\!C_\Phi\!=\!C_{\Phi^\pm}\!=\!3/2$ with $\Phi\!=\!h,H,A,\phi$ and $\Phi^\pm\!=\!H^\pm,\phi^\pm$.

The field-dependent squared masses, entering Eq.~(\ref{eq:CW}), are given by
\begin{equation}
\hat{m}_{V}^{2} \left( h, H \right) = \frac{h^{2}+H^{2}}{v^{2}} m_{V}^{2} \dist \hat{m}_{f}^{2} \left( h \right) = \frac{h^{2}}{2}y_{f}^{2}\,,
\end{equation}
for gauge bosons $V=W^{\pm}, Z$ and fermions, respectively, were $m_{V}$ are the masses at zero temperature, and $y_{f}$ the corresponding Yukawa couplings. The remaining terms are obtained as the eigenvalues of the scalar bosons' mass matrices
\begin{equation}
\label{eq:MS}
\begin{split}
\widehat{M}_{S}^{2} = & \frac{1}{2} \begin{pmatrix}
6\lambda_{1}h^{2} - 2\lambda_{1}v^{2} + \lambda_{345}H^{2} & 2hH\lambda_{345} \\ 
2hH\lambda_{345} & 6\lambda_{2}H^{2} + \lambda_{345}h^{2} + 2\mu_{2}^{2}
\end{pmatrix} \\
\widehat{M}_{P}^{2}  = & \frac{1}{2} \begin{pmatrix}
2\lambda_{1}h^{2} - 2\lambda_{1}v^{2} + \bar{\lambda}_{345}H^{2} & 2hH\lambda_{5}\\
2hH\lambda_{5} & 2\lambda_{2}H^{2} + \bar{\lambda}_{345}h^{2} + 2\mu_{2}^{2}
\end{pmatrix} \\
\widehat{M}_{\pm}^{2}  = & \frac{1}{2} \begin{pmatrix}
2\lambda_{1}h^{2} - 2\lambda_{1}v^{2} + \lambda_{3}H^{2} & hH\left( \lambda_{4} + \lambda_{5} \right) \\ 
hH\left( \lambda_{4} + \lambda_{5} \right) & 2\lambda_{2}H^{2} + \lambda_{3}h^{2} + 2\mu_{2}^{2}
\end{pmatrix} \ ,
\end{split}
\end{equation}
which coincide with (\ref{eq:MM}) in the EW ($T=0$) vacuum $(h,H)=(v,0)$.

To compensate shifts of the vacuum due to one-loop corrections, the counterterm potential
\begin{equation}
\Vct \left( h,H \right) = \delta m_{h}^{2} h^{2} + \delta m_{H}^{2} H^{2} + \delta \lambda_{1} h^{4}
\end{equation}
is assumed with coefficients following from the renormalization conditions
\begin{equation}
\label{eq:CT}
\begin{split}
\left. \frac{\partial \Vct}{\partial h} \right\vert_{\VEV} & = \left. -\frac{\partial \Vcw}{\partial h} \right\vert_{\VEV} \\
\left. \frac{\partial^{2} \Vct}{\partial h^{2}} \right\vert_{\VEV} & =- \left. \left(\frac{\partial^{2}\!\left.\Vcw\right\vert_{n_{\!\phi^{(\pm)}}=0}}{\partial h^{2}} + \frac{1}{32\pi^{2}} \sum_{i=\phi,\phi^{\pm}} n_{i} \left( \frac{\partial \hat{m}_{i}^{2}\left( h,H\right)}{\partial h} \right)^{2} \ln \frac{m_{\rm IR}^2}{Q^{2}} \right) \right\vert_{\VEV} \\
\left. \frac{\partial^{2} \Vct}{\partial H^{2}} \right\vert_{\VEV} & =- \left. \left(\frac{\partial^{2}\!\left.\Vcw\right\vert_{n_{\!\phi^{(\pm)}}=0}}{\partial H^{2}} + \frac{1}{32\pi^{2}} \sum_{i=\phi,\phi^{\pm}} n_{i} \left( \frac{\partial \hat{m}_{i}^{2}\left( h,H\right)}{\partial H} \right)^{2} \ln \frac{m_{\rm IR}^2}{Q^{2}} \right) \right\vert_{\VEV} \ .
\end{split}
\end{equation}
While this fixes the renormalization of the masses $m_{h,H}$ of the SM Higgs and the DM particle, as well as the SM coupling $\lambda_{1}$, the new couplings $\lambda_{2}$ and $\lambda_{345}$ are running $\overline{\rm MS}$ couplings and the masses $m_{A,H^{\pm}}$ are assumed to be one-loop corrected.
The second derivatives of $\Vcw$ contain in principle ill-defined terms, due to the inclusion of Goldstone modes that contribute at finite temperature but are massless at $T=0$ in the Landau gauge, leading to IR divergences (see, \emph{e.g.}, Ref.~\cite{Counterterm2HDM}). These expressions are subtracted, introducing an IR cutoff $m_{\mathrm{IR}}^2=m_{h}^{2}$.
This is realized in Eq.~(\ref{eq:CT}) above effectively by removing the Goldstone modes from the CW potential in the second and third line and adding instead the regular sums over Goldstone modes on the right-hand sides.

We now move on to include finite temperature effects, captured by the one-loop CW potential derived including thermal corrections to the n-point functions~\cite{Dolan_1974}, reading
\begin{equation}
\label{eq:VT}
V_{T} \left( h, H \right) = \frac{T^{4}}{2\pi^2} \left[ \sum_{i} n_{i}^{\mathrm{B}} J_{\mathrm{B}}\left( \frac{\tilde{m}_{i}^{2} \left( h,H,T \right)}{T^{2}} \right) + \sum_{i} n_{i}^{\mathrm{F}} J_{\mathrm{F}}\left( \frac{\hat{m}_{i}^{2} \left( h,H \right)}{T^{2}} \right) \right]\,,
\end{equation}
with the sums running over all viable bosons and fermions, respectively. The corresponding thermal functions are defined as~\cite{Anderson_1992}
\begin{equation}
J_{\mathrm{B}/\mathrm{F}} \left( x \right)  \defEq  \pm \int_{0}^{\infty} \dd t \ t^{2} \ln \left[ 1 \mp e^{-\sqrt{t^{2} + x}} \right] = \lim_{N\rightarrow \infty} \mp \sum_{l=1}^{N} \frac{\left( \pm 1 \right)^{l}x}{l^{2}}K_{2} \left( \sqrt{x}l \right)
\end{equation}
and can be well approximated by truncating the infinite sum at $N=5$, according to Ref.~\cite{PT2HDM}.

Since at the relevant temperatures of the phase transition, temperature-enhanced corrections spoil the convergence of the perturbative expansion, a resummation of such contributions is required. The problematic corrections can in fact be included by resumming the leading daisy self-energy diagrams, which is accounted for in Eq.~(\ref{eq:VT}) by dressing the field-dependent masses with thermal corrections \cite{Gross_1981,ParwaniApproach} $\hat{m}_{i}^{2} \left(h,H \right) \to \tilde{m}_{i}^{2} \left( h,H,T \right)$. The squared scalar mass matrices in (\ref{eq:MS}) now become $\widetilde{M}_{X}^{2} \equiv \widehat{M}_{X}^{2}+ \widehat{\Pi}(T),\, X\!=\!S,P,\pm$, with the $2\times 2$ diagonal matrix $\widehat{\Pi}$, derived from the low-energy limit of the respective two-point functions, with components
\begin{equation}
\begin{split}
\widehat{\Pi}_{11} \left( T \right) & =  \frac{T^{2}}{24} \left( 6y_{t}^{2} + 6y_{b}^{2} + 2y_{\tau}^{2} + \frac{9}{2}g_{W}^{2} + \frac{3}{2}g^{\prime 2} + 
12\lambda_{1} + 4\lambda_{3} + 2\lambda_{4} \right) \\
\widehat{\Pi}_{22} \left( T \right) & =  \frac{T^{2}}{24} \left( \frac{9}{2}g_{W}^{2} + \frac{3}{2}g^{\prime 2} + 12\lambda_{2} + 4\lambda_{3} + 2\lambda_{4} \right) \ .
\end{split}
\end{equation}
The transversal parts of the SM gauge bosons are not affected by finite-temperature corrections, whereas the squared Debye masses for the longitudinal components read \cite{PT2HDM}
\begin{equation}
\tilde{m}_{W_{L}}^{2} = \frac{h^{2}+H^{2}}{v^{2}}m_{W}^{2} + 2g_{W}^{2}T^{2},\
\ \tilde{m}_{Z_{L},\gamma_{L}}^{2} = \frac{h^{2}+H^{2}}{8} \left( g_{W}^{2} + g^{\prime 2} \right) + \left( g_{W}^{2} + g^{\prime 2} \right) T^{2} \pm \Delta \ ,
\end{equation}
and are weighted with $n_{W_L}\!=\!2,\,n_{Z_L}\!=\!n_{\gamma_L}\!=\!1,\ C_{W_L}\!=\!C_{Z_L}\!=\!3/2,\,C_{\gamma_L}\!=\!0$, with the squared substitution 
\begin{equation}
\Delta^{2} \defEq \frac{\left( h^{2} + H^{2} + 8T^{2} \right)^{2}}{64} \left( g_{W}^{2} + g^{\prime 2} \right)^{2} - g_{W}^{2}g^{\prime 2}T^{2} \left( h^{2} + H^{2} + 4T^{2} \right) \ .
\end{equation}

In a one-step first-order EWPhT, $\braket{h}$ departs from zero at the critical temperature $T_{c}$ at which the minimum at the origin and the minimum at $v_{c} > 0$ are degenerate. The strength~$\xi$ of such EWPhT can be estimated via the critical temperature and the corresponding field value as
\begin{equation}
\xi \defEq \frac{v_{c}}{T_{c}} \ .
\end{equation}
Allowing for an intermediate departure of $\braket{H}$ from zero leads to a two-step EWPhT. This departure takes place at a temperature that we will denote by $T_{2}$ and the departure of $\braket{h}$ from zero occurs in turn at a temperature $T_{1}< T_{2}$. Thus, it is possible to define two EWPhT strengths $\xi_{1,2}$ accounting for the latter and former transition, respectively, which we take as 
\begin{equation}
\xi_{j} \defEq \frac{\sqrt{\braket{h}_{j}^{2} + \braket{H}_{j}^{2}}}{T_{j}}
\label{Eq:EWPTstrength2step}
\end{equation}
with $j = 1,2$ and the corresponding vevs $\braket{h}_j,\braket{H}_j$ eventually emerge at the transition temperatures $T_{j}$. 
Note that the definition of the EWPhT strength in Eq.~(\ref{Eq:EWPTstrength2step}) also holds in the case of a one-step EWPhT. Although the correct criterion %to prevent the washout due to sphalerons 
for a strong EWPhT in the context of an additional doublet as well as its gauge dependence is not comprehensively clarified yet, we assume such an EWPhT capable to allow for EW baryogenesis to be present for $ \xi_{j} \geq 1$. 
While different definitions of $\xi_j$, compared to Eq.~(\ref{Eq:EWPTstrength2step}), could be envisaged for a two-step transition, a crucial quantity for phase transitions in the context of EW baryogenesis is the EW sphaleron rate. These sphalerons are in fact sufficiently suppressed for $ \xi_{j} \gtrsim 1$. Still, the question of defining the best-suited criterion for multiple-step EWPhT invites for further investigation.

\section{Dark Matter Physics}
\label{sec:DM}

\begin{figure*}[b!]
\centering
\includegraphics[width=0.9\textwidth]{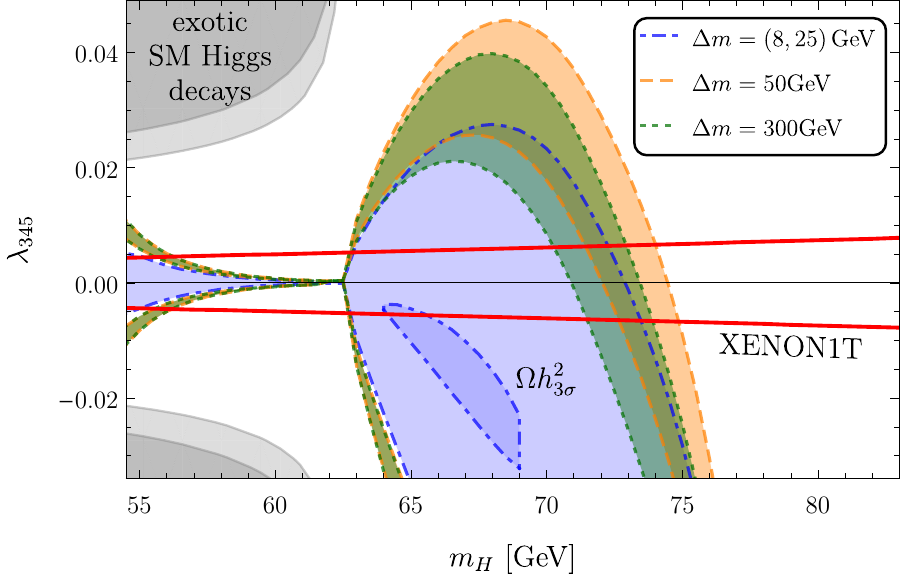}
\caption{Contours of relic abundance within $0.6\Omega h^{2}_{\mathrm{best}}$ (outer lines) and $\Omega h^{2}_{\mathrm{best}}\! +\! 3 \sigma$ (inner lines) for different mass splittings $\Delta m \equiv m_{A,H^{\pm}}\!-\!m_{H}$. For the smallest splitting $\Delta m = (8,25)\GeV= (m_{A}-m_{H},m_{H^{\pm}}-m_{H})$, the inner contour encloses the region with an abundance of at least $\Omega h^{2}_{3\,\sigma} \equiv \Omega h^{2}_{\mathrm{best}} - 3 \sigma$. Constraints from the absence of exotic SM Higgs decays for $m_{H} < m_{h}/2$ and the recent XENON1T results, indicated by the red line, reduce the viable parameter space.}
\label{Fig:DMConstraints}
\end{figure*}
The IDM naturally features a promising DM candidate, represented by the lightest $\mathbb{Z}_2$-odd scalar, given that it is electromagnetically neutral. In the following we will assume that this role is played by the CP even state $H$. In the light of DM phenomenology (and EWPhT in the subsequent analysis), the exchange $H \leftrightarrow A$ for the DM particle along with $\lambda_{345} \leftrightarrow \bar{\lambda}_{345}$ gives rise to the same results (see, \emph{e.g.}, Ref.~\cite{IDM}). As it will turn out, in relevant regions of the parameter space co-annihilations of $H$ with other states will need to be considered in calculating the relic-abundance.\footnote{Such co-annihilations can appear quite naturally in models with additional weak multiplets.} This can be straightforwardly included into the Boltzmann equation, describing
the evolution of the DM number density $n_H$, \cite{GondoloBoltzmannEquation,CosmicAbundances} 
\begin{equation}
\frac{\dd n_H}{\dd t} = -3Hn_H -\braket{\sigma_{\mathrm{eff}}v} \left[ n_H^{2} - \left( n_H^{\mathrm{eq}}\right)^{2} \right]\,,
\end{equation}
employing the effective thermally averaged annihilation cross section times velocity
\begin{equation}
\braket{\sigma_{\mathrm{eff}}v} \defEq \sum_{j=1}^{N} \braket{\sigma v}_{Hj} \frac{n_{H}^{\mathrm{eq}} n_{j}^{\mathrm{eq}}}{\left( n^{\mathrm{eq}} \right)^{2}} \ ,
\end{equation}
which takes co-annihilations into account. Here, $j$ runs over all $N$ potential co-annihilating states. Assuming a Maxwell-Boltzmann distribution, the particle density $n^{\mathrm{eq}}$ in thermal equilibrium is given by
\begin{equation}
n^{\mathrm{eq}} = \sum_{i} n_{i}^{\mathrm{eq}}= \frac{T}{2\pi^{2}} \sum_{i} g_{i}m_{i}^{2} K_{2} \left( \frac{m_{i}}{T} \right)\,, \label{Eq:EquiDensity}
\end{equation}
with the temperature $T$ and the mass $m_{i}$ and degrees of freedom $g_{i}$ of the non-SM particles. 
For the computation of the relic abundance $\Omega h^{2}$ (and the DD cross section), we employ the package micrOMEGAs 5.0.8~\cite{micromegasManual} to solve the Boltzmann Equation, together with the CalcHEP package~\cite{calchep}, delivering also the bare cross sections $\sigma$ as well as the thermally averaged annihilation cross sections.
Below, we will use the fit $\Omega h^{2}= 0.1200(12)$ for the relic abundance with the reduced Hubble expansion rate $h = 0.674(5)$ from Ref.~\cite{PDG}, denoting the central value as $\Omega h^{2}_{\mathrm{best}}$. 

The resulting viable slices of parameter space for three mass spectra, defined by the splittings $\Delta m \equiv m_{A,H^{\pm}}-m_{H}$, that neither violate any of the theoretical constraints discussed in Sec.~\ref{sec:ThBounds}, nor exceed the measured relic abundance, are shown in Fig.~\ref{Fig:DMConstraints}.
From the plot, one can identify three particularly interesting regions
(see also, \emph{e.g.}, Refs.~\cite{IDM,Dolle:2009fn,Profumo}):\\
\\
\textbf{Funnel region}: For DM masses $m_{H}<m_{h}/2$, the process $HH \rightarrow h^{*} \rightarrow b\bar{b}$ dominates, mediated by an off-shell SM Higgs boson
and governed by the $hHH$ interaction strength $\sim v \lambda_{345}$. Thus, $\Omega h^{2} \propto \left\vert \lambda_{345} \right\vert^{-2}$. Since the annihilation cross section increases by approaching the resonance pole at half of the SM Higgs mass, the relic abundance decreases. The curve for the full relic abundance is symmetric with respect to zero coupling.\\
\\
\textbf{Resonance region}: The annihilation of two DM particles into SM particles via an on-shell SM Higgs boson is favoured at half the SM Higgs boson mass $m_{h}/2 \approx 62.5 \GeV$, resulting in a large, resonantly enhanced, cross section. That leads to a relic abundance comparable to the measured value for very small portal couplings only.\\
\\
\textbf{'Tail' region}: For DM masses $m_{H} \gtrsim m_{h}/2$, contributions from annihilation processes featuring (off-shell) gauge bosons become significant (we will comment on co-annihilations via intermediate weak bosons further below). Consequently, the amplitude is not simply proportional to $\lambda_{345}$ any more and interference effects occur. The closer the DM mass gets to the gauge boson mass, the stronger the contribution. In consequence, the coupling parameter $\lambda_{345}$ needs to become negative to cancel the contributions of additional annihilation channels. The process $HH \rightarrow W^{+}f\bar{f^{\prime}}$, with an off-shell $W^{-}$-boson decaying into a fermion-antifermion pair (\emph{e.g.}, $W^{-*} \rightarrow e^{-}\bar{\nu}_{e}$), occurs either via a four-point interaction with an SM gauge-coupling strength or via $s$-, $t$- or $u$-channel with an intermediate SM Higgs $h$ or a $H^{-}$, respectively. The same holds for $HH \rightarrow Z^\ast Z$ with possible $A$-mediated channels.\\
\\
Imposing constraints from searches for exotic SM Higgs decays and from the XENON1T experiment, the surviving parameter space resides in the DM mass range $55\GeV \lesssim m_{H} \lesssim 75\GeV$ and in the Higgs portal coupling range $\left\vert \lambda_{345} \right\vert \lesssim 0.01$, with the latest DD limits removing significant portions of parameter space. 

A comment is in order regarding the small splitting scenario, where we considered two splittings of $m_{A}-m_{H}= 8\GeV$, not to violate Eq.~(\ref{eq:LEPsplit}), and $m_{H^\pm}-m_{H}= 25\GeV$. The sizable second splitting suppresses co-annihilations, discussed in detail below, that otherwise would lead to significantly under-abundant DM. This allows for a new scenario for light DM at $m_H\gtrsim m_h/2$ in the IDM, where a fair DM abundance is not limited to small tuned regions (see also \cite{Dolle:2009fn,Lu:2019lok}).

Beyond that, an additional parameter-space region with the correct relic abundance opens in the high-mass regime $m_{H} \gtrsim 600\GeV$, in particular for small mass splittings $\Delta m \lesssim 10\GeV$. The latter tendency can be understood, noting that the amplitudes for the crucial annihilation channels into longitudinal $W$ and $Z$ bosons scale as $\lambda_3$ and $\bar \lambda_{345}$, respectively. Inspecting Eqs.~(\ref{eq:lamdef}) and (\ref{eq:lambdas}), it becomes evident that both of these couplings can become small for small $\Delta m$ (given that $\lambda_{345}$ is not too large). This allows for a suppressed annihilation and in turn for an appreciable DM abundance.
A detailed study of this mass regime in the context of DM can be found in, \emph{e.g.}, Ref.~\cite{IDM}, and we will add the connection with the emerging EWPhT towards the end of this article. We also note that the presence of the $D=6$ operator ${\cal O}_{H_2 \tilde F}$ from Eq.~(\ref{eq:CPo}) would have an impact on that region, modifying the annihilation of DM into $W$ bosons. It would be interesting to include this effective operator, with a proper coefficient inducing sufficient CP violation to make baryogenesis fully viable, together with further potentially generated $D=6$ operators in the analysis and study their impact on the relic abundance. This would, however, require significant changes in the analysis tools and is thus left for future work. 

\begin{figure*}[b!]
\centering
\begin{minipage}{0.49\textwidth}
\centering
\includegraphics[width=\textwidth]{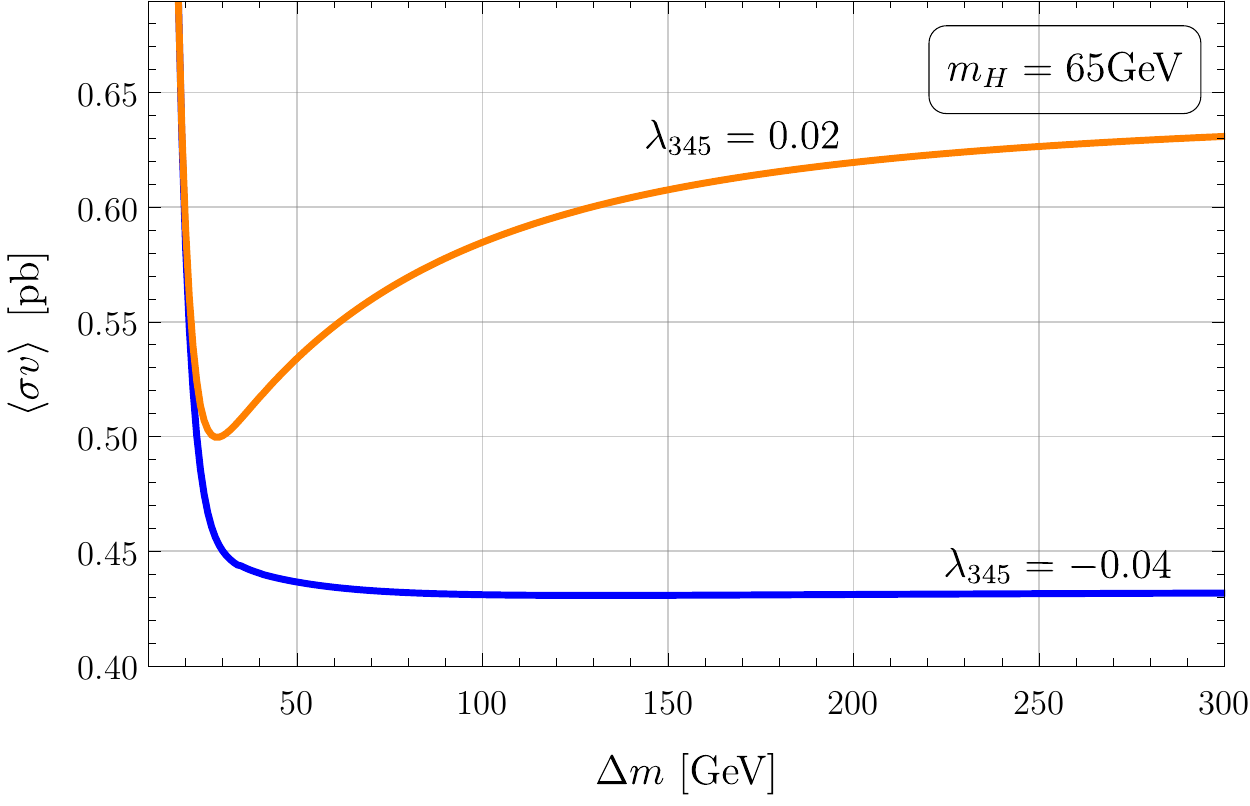}
\end{minipage}\\
\begin{minipage}{0.49\textwidth}
\centering
\includegraphics[width=\textwidth]{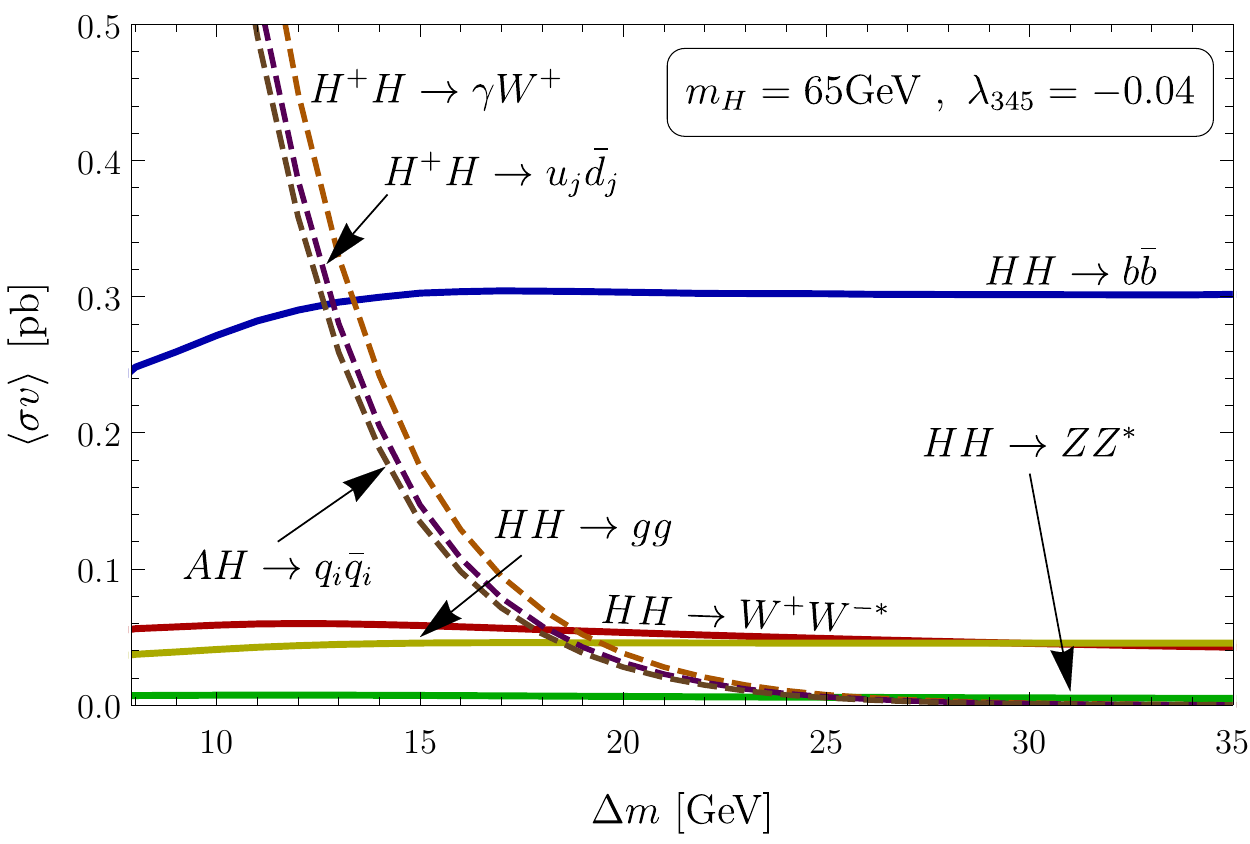}
\end{minipage}
\begin{minipage}{0.49\textwidth}
\centering
\includegraphics[width=\textwidth]{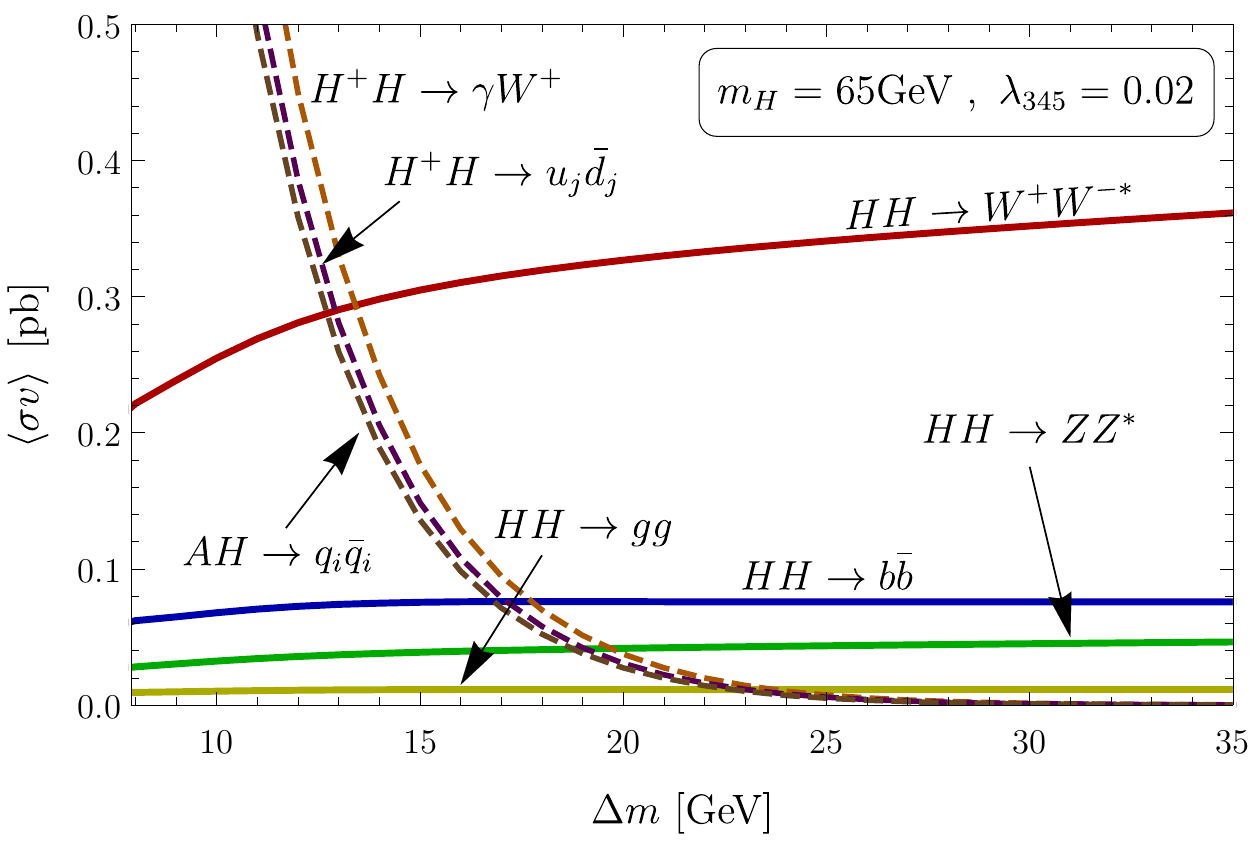}
\end{minipage}
\caption{Dependence of the thermally averaged cross section on the mass splitting $\Delta m$ for fixed DM mass $m_{H}=65\,$GeV and two Higgs portal couplings, $\lambda_{345}=-0.04,0.02$. \textsc{Upper row}: Total \mbox{(co-)annihilation} cross sections.   
\textsc{Lower row}: Comparison of (co-)annihilation channels with $j(i) = 1,2(,3,4,5)$. At small mass splitting, co-annihilations (dashed) into $\gamma W^{+}$ and quarks 
dominate, while the annihilation channels (solid) dominate for large $\Delta m$.}
\label{Fig:ChannelAnalysis}
\end{figure*}

In Fig.~\ref{Fig:ChannelAnalysis}, the behavior of the thermally averaged cross section depending on the mass splitting $\Delta m$ is analyzed for a fixed DM mass of $m_{H}=65\GeV$ and two values of the portal parameter, $\lambda_{345}=-0.04$ and $\lambda_{345}=0.02$.
The upper panel shows the total cross section, which is largely enhanced due to significant co-annihilations of a DM state $H$ and another non-SM state $A$, $H^{\pm}$ for small mass splittings, allowing for simultaneous non-negligible presence of those $\mathbb{Z}_2$-odd states around freeze-out. This behavior is similar for both portal-coupling values, where around $\Delta m \approx 30\GeV$ the steep decrease ends. For sizable splittings, $\braket{\sigma v}$ approaches quickly a constant value for $\lambda_{345}=-0.04$, whereas for $\lambda_{345}=0.02$ it first starts to increase and reaches a higher plateau for large $\Delta m$. Indeed, the contributions from co-annihilation channels become irrelevant for increased mass splitting, as can be seen in Eq.~(\ref{Eq:EquiDensity}) where a large splitting results in a small fraction of the non-DM particle in the particle density at thermal equilibrium. 

The relative differences between the various co-annihilation processes are dissected in the lower panel of Fig.~\ref{Fig:ChannelAnalysis}. They are determined by the mass of the potential mediator and the coupling of the SM boson. The same is true for the DM annihilation into a $b\bar{b}$-pair via an intermediate SM Higgs boson. In addition, the pair creation of SM gauge bosons $W^{\pm}$, $Z$ can occur via a four-point interaction. One can conclude from the plots in the lower row that co-annihilation processes can become important in the considered DM mass range for small splittings. In fact, for small overall $\Delta m \lesssim 10\,$GeV, they would lead unavoidably to underabundance  in the intermediate mass range of $m_H\sim 65-75$\,GeV. However, allowing for somewhat heavier $H^\pm$ while keeping $m_A$ close to $m_H$ can suppress the more dangerous $H^\pm$ co-annihilations sufficiently to allow for a low-splitting dark matter scenario in agreement with LEP constraints and achieving the full DM abundance, as envisaged before, in the context of Fig.~\ref{Fig:DMConstraints}.

\section{Scanning the Nature of the Electroweak Phase Transitions}
\label{sec:EWPhT}

\begin{figure*}[b!]
\centering
\includegraphics[width=0.99\textwidth]{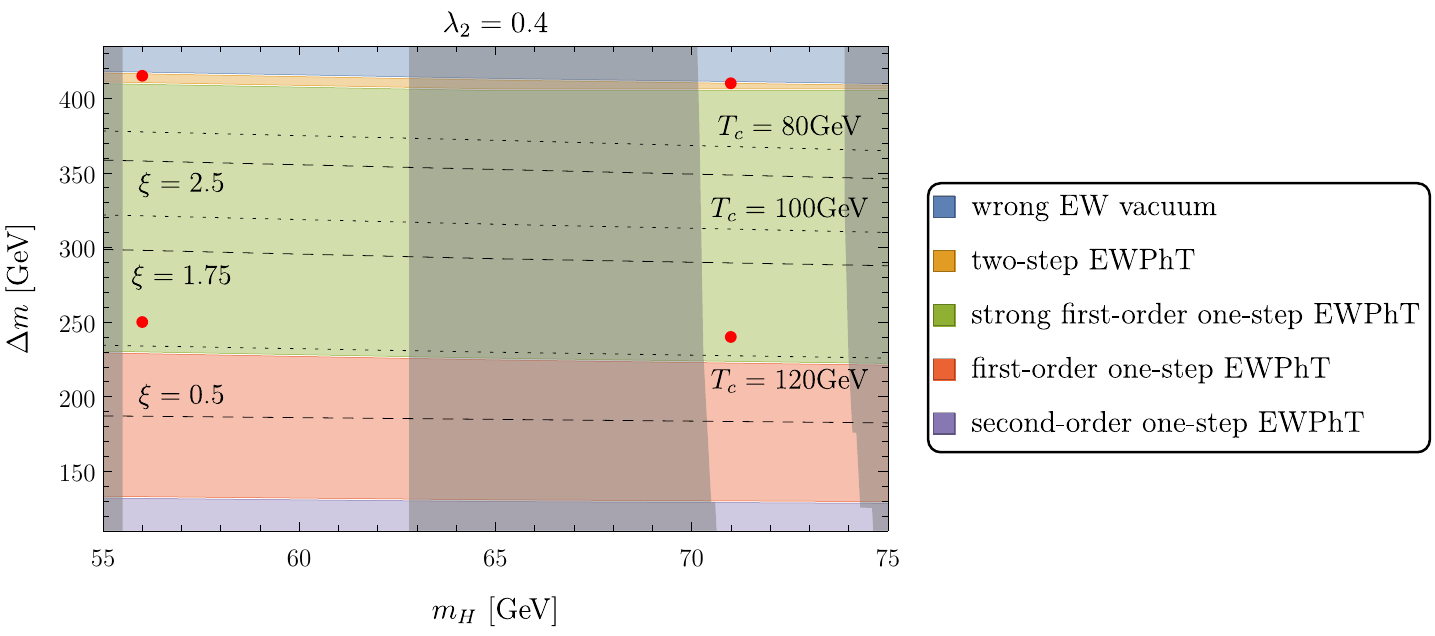}
\caption{Characterization of the EWPhT (colored stripes), indicating the surviving parameter space (bright regions) which provides a significant amount of the measured DM abundance, \emph{i.e.}, $0.6\Omega h^{2}_{\mathrm{best}} \leq \Omega h^{2} \leq \Omega h^{2}_{\mathrm{best}} +3\sigma$, and obeys the latest DD limits from XENON1T, adjusting $\lambda_{345}\lesssim 0.005$. %, employing $\lambda_{345} = 0.005$. 
The red points denote the selected BMs (see Tab.~\ref{Tab:BMPsOverview}), while dashed and dotted contours show the EWPhT strength $\xi$ and the critical temperature $T_{c}$, respectively. See text for details.}
\label{Fig:EWPT_DM_BMPs}
\end{figure*}

After having identified viable parameter-space regions for DM, the goal of this section is to investigate different kinds of EWPhT emerging in the surviving parameter space. Considering the constraints from the relic abundance and DD limits from Fig.~\ref{Fig:DMConstraints}, we first focus on the low DM-mass range $55 \GeV \leq m_{H} \leq 75 \GeV$ -- the high-mass regime will be explored further below. The dependence of the type of EWPhT on the degenerate masses $m_{A}=m_{H^{\pm}}$ of the $\mathbb{Z}_{2}$-odd non-DM particles is visualised in Fig.~\ref{Fig:EWPT_DM_BMPs} for  $\lambda_{2}=0.4$. 
Considering DM physics, the Higgs portal coupling $\lambda_{345}$ is adapted to obtain the measured relic abundance in agreement with XENON1T limits. Regions where this is not possible and the correct abundance cannot be obtained, given DD constraints, are shaded in gray (\emph{cf.} Fig.~\ref{Fig:DMConstraints}).
Since the type of EWPhT is rather insensitive to $\lambda_{345}$ in the considered ranges of $\lambda_{345}\lesssim 0.005$ and $55\,{\rm GeV} \leq m_{H} \leq 75\,{\rm GeV}$, we used a fixed $\lambda_{345}=0.005$ for the EWPhT part of the plot.
The different kinds of transition are depicted by different colors, where the two-step EWPhT is strongly first-order implicitly. Finally, the critical temperature $T_{c}$ and the strength of the EWPhT $\xi$ are indicated by dotted and dashed contours, respectively.

A large range in $m_{A,H^{\pm}}$ with a strong first-order one-step EWPhT, \emph{i.e.}, $\xi\geq 1$, is present (green region), while a narrow mass range allows a viable strong two-step EWPhT, not conflicting other constraints (yellow stripe).\footnote{Although the IDM has been studied already in this context (see, \emph{e.g.}, Refs.~\cite{Ginzburg_2010,Gil:2012ya,Blinov_2015}), this region where a two-step EWPhT is in agreement with DM phenomenology was not found yet. See also Refs.~\cite{Land_1992,Zarikas_1996,Inoue_2016,Aoki_2021} for analyses of two-step transitions in a generic Two-Higgs-Doublet model.} The presented fine-grained numerical treatment enables to resolve this small range of mass splitting.
Varying both $\lambda_{2}$ and $\lambda_{345}$, we still found that only a small window remains for the future exploration of the two-step EWPhT, regardless of the coupling parameters above. Finally, there is an upper bound for the mass splitting, beyond which the EW vacuum at zero temperature is no longer at $\left( v, 0 \right) = \left( 246, 0 \right)\GeV$, as indicated by the light-blue region.
%The latter had not been found before in the literature, since no (dense) scans in the corresponding region were performed (see, \emph{e.g.}, Ref.~\cite{Gil:2012ya}). See also Refs.~\cite{Land_1992,Zarikas_1996,Inoue_2016,Aoki_2021} for an analysis of two-step transitions in a generic Two-Higgs-Doublet model.

From Fig.~\ref{Fig:EWPT_DM_BMPs} we choose four benchmarks (BMs), two of which lead to a strong first-order EWPhT and the other two exhibit a two-step EWPhT. The mass spectra together with a proper coupling parameter $\lambda_{345}$, inducing the correct relic abundance while eluding XENON1T DD limits, are given in Tab.~\ref{Tab:BMPsOverview}. The small portal couplings for the BMs are due to the strong upper limits on $\sigma_{n,p}$, which can only be obeyed by tiny $\left\vert \lambda_{345} \right\vert$. In particular the BMs with two-step transitions will be promising targets for an analysis of GW signatures in the future, in the spirit of Ref.~\cite{Morais:2019fnm} (see also Refs.~\cite{Paul:2019pgt,Barman:2019oda,Borah:2020wut}).

\begin{table}[t!]
\centering
\begin{tabular}{ccccccc}
\hline 
$\mathrm{BM}$ & $m_{H} \ \left[ \mathrm{GeV} \right]$ & $m_{A,H^{\pm}} \ \left[ \mathrm{GeV} \right]$ & $\lambda_{345}$ & $\Omega h^{2}$ & $\sigma_{n} \ \left[ 10^{-13} \mathrm{pb} \right]$ & $\sigma_{p} \ \left[ 10^{-13} \mathrm{pb} \right]$ \\ 
\hline 
$1$ & $56$ & $306$ & $0.0037$ & $0.1188$ & $379.7$ & $372.2$ \\ 
$2$ & $56$ & $471$ & $0.0037$ & $0.1188$ & $379.7$ & $372.2$ \\ 
$3$ & $71$ & $311$ & $0.0020$ & $0.1210$ & $69.5$ & $68.1$ \\ 
$4$ & $71$ & $481$ & $0.0020$ & $0.1177$ & $69.5$ & $68.1$ \\ 
\hline 
\end{tabular}
\caption{Benchmark points for the further investigation of the EWPhT dynamics. The relic abundance is within $3\sigma$ around $\Omega h^{2}_{\mathrm{best}}$ and the cross sections $\sigma_{n,p}$ for scattering off a neutron or a proton, respectively, obey the latest XENON1T constraints.}
\label{Tab:BMPsOverview}
\end{table}

\begin{figure}[b!]
\centering
\begin{minipage}{0.49\textwidth}
\centering
\includegraphics[width=\textwidth]{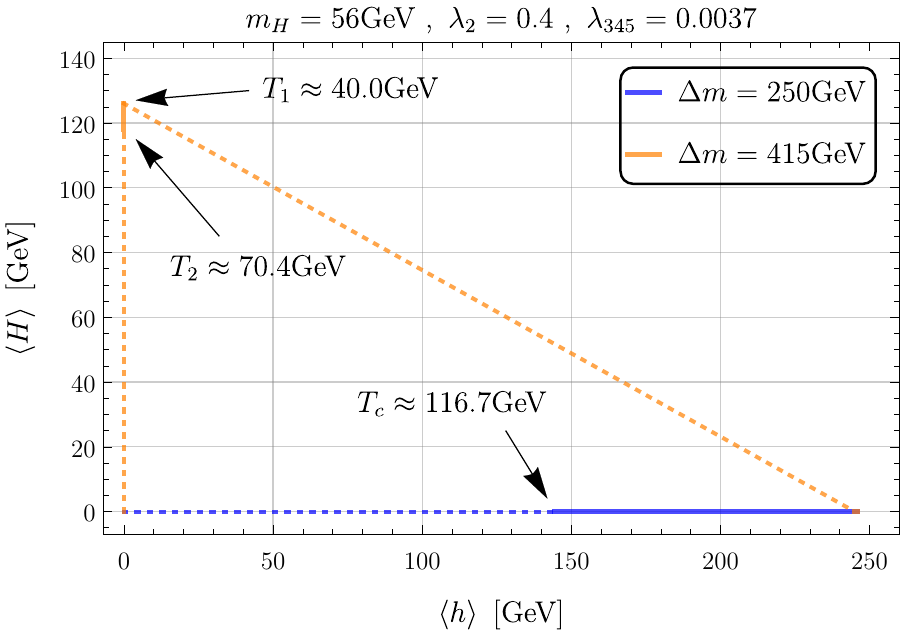}
\end{minipage}
\begin{minipage}{0.49\textwidth}
\centering
\includegraphics[width=\textwidth]{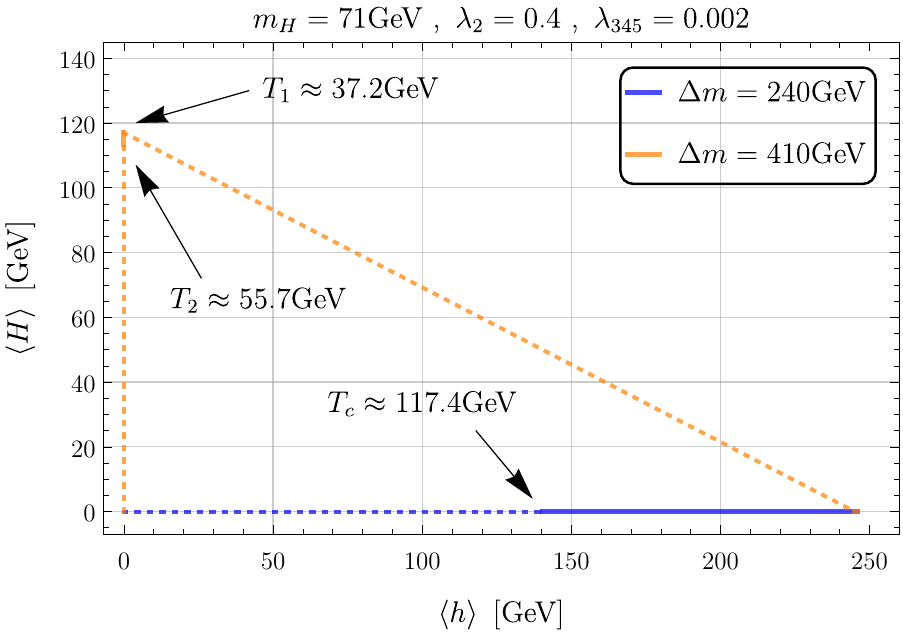}
\end{minipage}
\caption{Evolution of the fields for the BMs. Solid lines correspond to a smooth evolution of the fields values, while dashed lines indicate a non-continuous transition. For the one-step EWPhT, the transition occurs around the critical temperature, denoted by $T_{c}$, while for the two-step EWPhT
there are two temperatures, $T_{2}$ and $T_{1}$, for the first and second transition, respectively.}
\label{Fig:FieldFieldPlot}
\end{figure}

The evolution of the SM Higgs field $h$ and of the non-SM field~$H$ as the universe cools down is visualized in Fig.~\ref{Fig:FieldFieldPlot} for the four BMs in Tab.~\ref{Tab:BMPsOverview}, showing the characteristic trajectories for one-step (blue) and two-step (yellow) EWPhTs.
Solid lines correspond to a smooth evolution of the fields, while dashed lines indicate a non-continuous step from the initial to the final point.
In the case of a one-step EWPhT, the scalar field $H$ does not acquire a vev different from zero, which is thus given by a straight line. Starting in the symmetric field configuration at the origin for sufficiently high temperatures, here the transition occurs at $T=T_{c}$. For $T<T_{c}$, the vacuum continues to evolve continuously towards the correct EW vacuum at $\left( v, 0 \right)$. For a two-step EWPhT, $\braket{H}$ is non-zero within a certain temperature interval, leading to a departure from a straight line. In this scenario, the first transition happens at $T=T_{2}$. After a continuous evolution of the field the second transition takes place at $T=T_{1}<T_{2}$ before the vacuum evolves continuously towards the correct one. For this, it is necessary to distinguish between two temperature-dependent strengths $\xi_{1} \left( T_{1} \right)$ and $\xi_{2} \left( T_{2} \right)$ in the two-step EWPhT case, see Eq.~(\ref{Eq:EWPTstrength2step}). The type of the EWPhT and the corresponding EWPhT strengths $\xi_{j}$ for the BMs are summarized in Tab.~\ref{Tab:BMPsPTstrength}.
\begin{figure}[b!]
\centering
\begin{minipage}{0.49\textwidth}
\centering
\includegraphics[width=\textwidth]{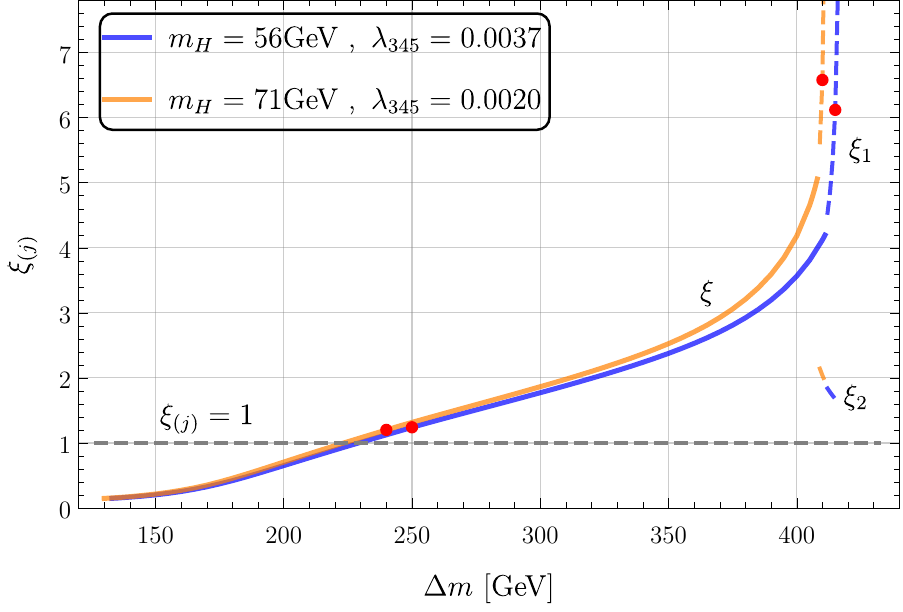}
\end{minipage}
\begin{minipage}{0.49\textwidth}
\centering
\includegraphics[width=\textwidth]{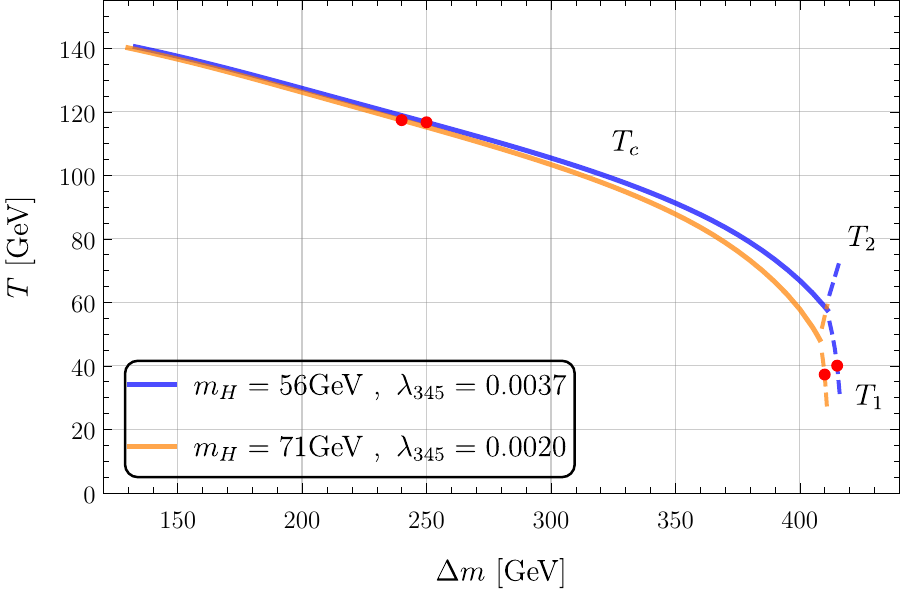}
\end{minipage}
\caption{Evolution of the EWPhT strengths $\xi_{j}$ and temperatures for different mass splittings $\Delta m$ with $\lambda_{2}=0.4$. Solid lines correspond to first-order EWPhTs whereas dashed lines to two-step EWPhTs. The EWPhT for $\Delta m \lesssim 130\GeV$ is of second order and for $\Delta m \gtrsim 420\GeV$ the EW vacuum does not match the proper $\left( v, 0 \right) \approx \left( 246, 0 \right)\,{\rm GeV}$. The benchmarks from Tab.~\ref{Tab:BMPsPTstrength} are depicted by red dots.}
\label{Fig:XiTcMassSplitting}
\end{figure}
The strength $\xi_j$ depends both on the mass splitting $\Delta m$ and on the parameter $\lambda_{2}$ that is not relevant in the DM analysis as we consider tree-level processes only. Thus, it is interesting to study the impact of these two quantities on the EWPhT. In the further analysis, we use the DM mass $m_{H}$ and the coupling parameter $\lambda_{345}$ of the BMs to examine the dependence of the strength $\xi$ on $\Delta m$ and $\lambda_{2}$.

In Fig.~\ref{Fig:XiTcMassSplitting}, the EWPhT strength $\xi_{j}$ both for one-step and for two-step EWPhT as well as the transition temperatures are shown. In accordance with Fig.~\ref{Fig:EWPT_DM_BMPs}, the EWPhT strength grows for increasing mass splitting. This behaviour holds both for one-step and for the second transition in the two-step EWPhT. The reason for that is the increase of the critical field value when the transition happens, together with the decrease of critical temperature, \emph{i.e.}, the EWPhT occurs at later times. Although the range of mass splittings in which a two-step EWPhT occurs is rather small, $\xi_{1}$ increases quickly, resulting in a large $\xi_{1}$ for the considered BMs. The EWPhT strength $\xi_{2}$, corresponding to the first transition where $\braket{H}\neq 0$, is smaller than $\xi_{1}$ due to the different transition temperatures. 
\begin{table}[t!]
\begin{tabular}{ccccccc}
\hline 
$\mathrm{BM}$ & $m_{H} \ \left[ \mathrm{GeV} \right]$ & $m_{A,H^{\pm}} \ \left[ \mathrm{GeV} \right]$ & $\lambda_{345}$ & $\mathrm{EWPhT \ type}$ & $\xi_{\left( 1 \right)}$ & $\xi_{2}$ \\ 
\hline 
$1$ & $56$ & $306$ & $0.0037$ & $\mathrm{strong \ first}\text{-}\mathrm{order \ one}\text{-}\mathrm{step}$ & $1.24$ & $-$ \\ 
$2$ & $56$ & $471$ & $0.0037$ & $\mathrm{two}\text{-}\mathrm{step}$ & $6.12$ & $1.68$  \\
$3$ & $71$ & $311$ & $0.0020$ & $\mathrm{strong \ first}\text{-}\mathrm{order \ one}\text{-}\mathrm{step}$ & $1.20$ & $-$ \\ 
$4$ & $71$ & $481$ & $0.0020$ & $\mathrm{two}\text{-}\mathrm{step}$ & $6.58$ & $2.04$ \\ 
\hline 
\end{tabular}
\centering
\caption{Type and strength of the EWPhT for the four BMs, assuming $\lambda_{2} = 0.4$.}
\label{Tab:BMPsPTstrength}
\end{table}

\begin{figure}[b!]
\centering
\begin{minipage}{0.49\textwidth}
\centering
\includegraphics[width=\textwidth]{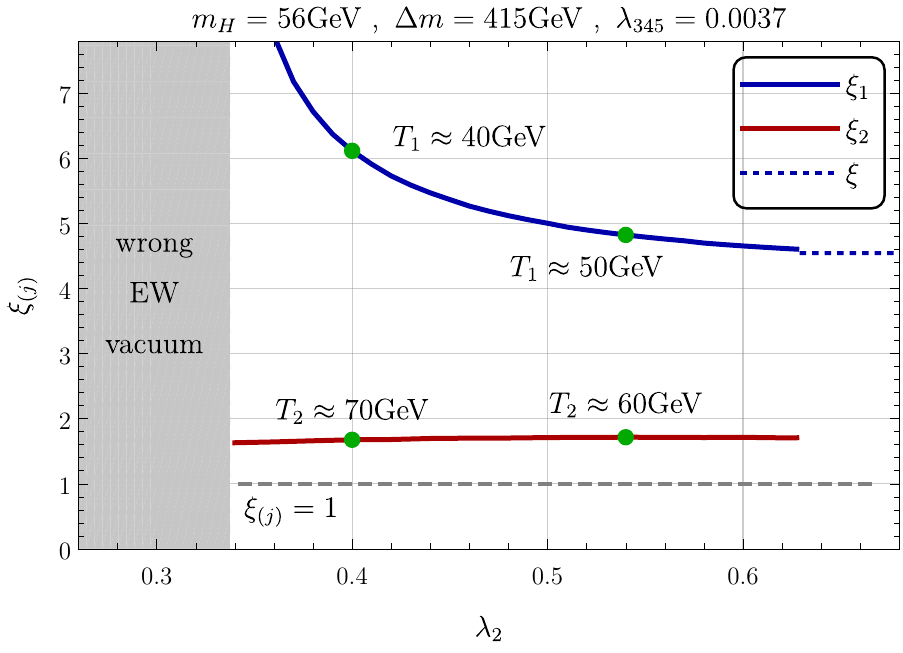}
\end{minipage}
\begin{minipage}{0.49\textwidth}
\centering
\includegraphics[width=\textwidth]{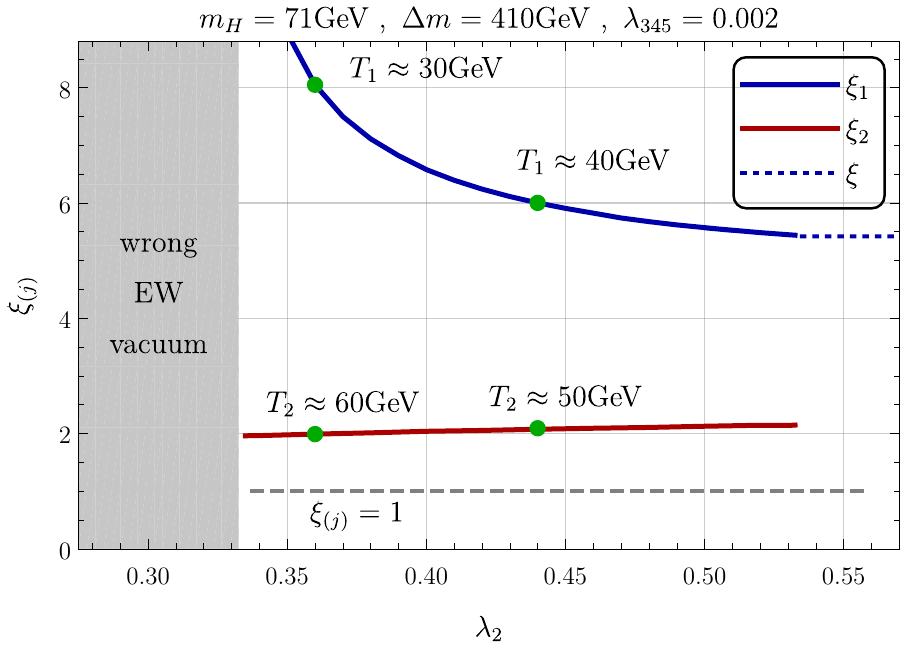}
\end{minipage}
\caption{Dependence of $\xi_{j}$ on the coupling parameter $\lambda_{2}$ for two BMs which represent two-step EWPhTs. The red lines correspond to the EWPhT strength of the first transition, whereas the blue lines to the later one.}
\label{Fig:XiLam2Dependence}
\end{figure}

The dependence of the two-step $\xi_{j}$ on the coupling $\lambda_{2}$ is shown in Fig.~\ref{Fig:XiLam2Dependence} for the two corresponding BMs, together with transition-temperature information. The parameter $\lambda_{2}$ is not completely free, as it is constrained from below by the EW vacuum condition. For sufficiently small $\lambda_{2}$, the global potential minimum is not located at $\left( v,0 \right)$ and is thus excluded. The exact value depends on the considered parameter set. Small values of $\lambda_{2}$ are favored to move from a  one-step to a two-step EWPhT via large thermal corrections. We note that very large values of $\xi_{1} \gtrsim 5$ are linked to rather low transition temperatures of $T_{1} \lesssim 50\GeV$.

Before concluding, we go back to the high DM-mass regime, $m_H \gtrsim 600\GeV$.
Here we expect it to be more difficult to simultaneously 
achieve a (strong) first-order EWPhT together with an appreciable DM abundance. The reason it that, as explained before in Sec.~\ref{sec:DM}, in this regime for significant mass splittings, \emph{i.e.}, sizable $\lambda_3$ and $\bar \lambda_{345}$, the annihilation into longitudinal $W$ and $Z$ bosons is large and the relic abundance is expected to be negligible. On the other hand this is the parameter space that is most promising for a strong EWPhT (\emph{cf.}\ Fig.~\ref{Fig:EWPT_DM_BMPs}). Still, this expectation is to be checked quantitatively (also because non-trivial cancellation could appear) and even a modest contribution of $H$ to the DM could be interesting for prospects of DD experiments, depending on the model parameters. Like this, a potential agent of electroweak baryogenesis could still be seen in DM searches.

\begin{figure*}[t!]
\centering
\includegraphics[width=0.99\textwidth]{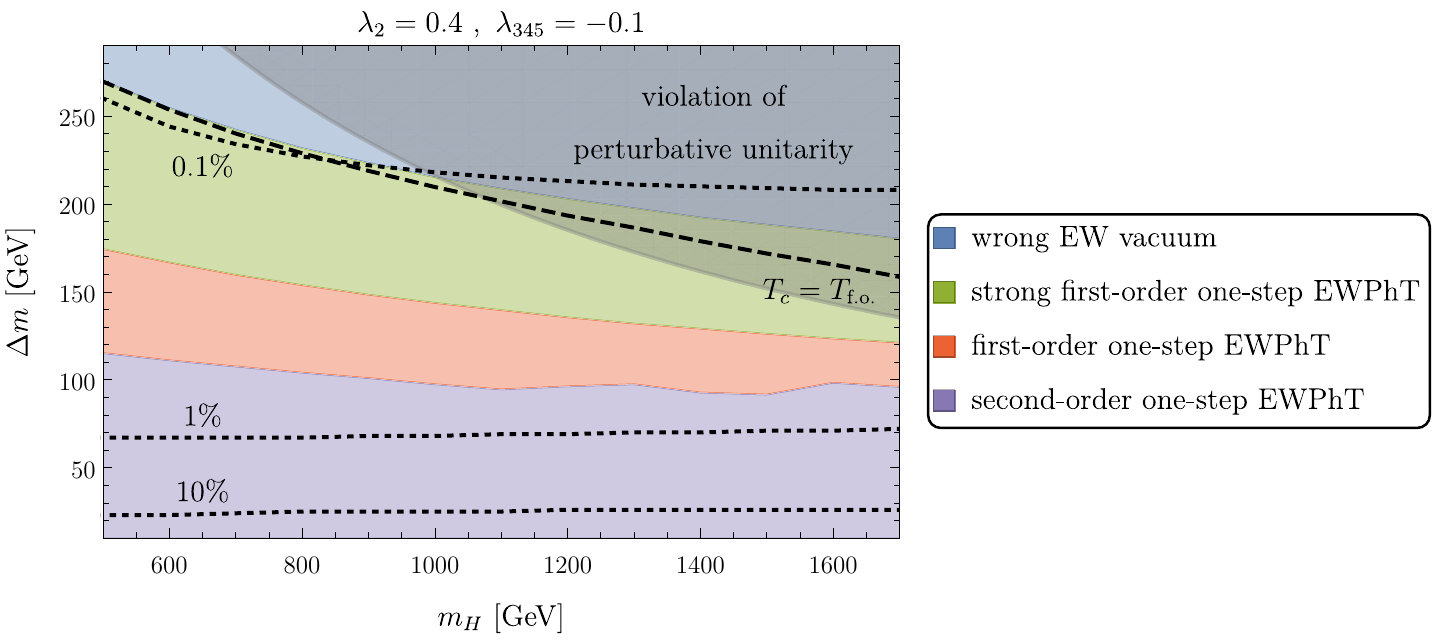}
\caption{Characterization of the EWPhT (colored stripes)
for the high-mass regime. The dotted contours indicate the fraction of the full relic abundance $\Omega h^{2}_{\mathrm{best}}$ that resides in $H$ and the dashed one shows the critical temperature $T_{c}$ being equal to the freeze-out temperature $T_{\mathrm{f.o.}}\approx m_{H}/20$. See text for details.}
\label{Fig:EWPT_DM_high}
\end{figure*}

Thus, in Fig.~\ref{Fig:EWPT_DM_high}, we present a similar characterization of the interplay of the EWPhT and DM physics as in Fig.~\ref{Fig:EWPT_DM_BMPs}, but now for heavy DM. While the color code remains the same as before, we now indicate, via dotted contours, the fraction of $\Omega h^{2}_{\mathrm{best}}$, assuming $\lambda_{2}=0.4$ and $\lambda_{345}=-0.1$ being well below the XENON1T bound. 
One can inspect that achieving a large fraction of the DM abundance would only be possible in the case of a second-order transition. On the other hand, a strong first order EWPhT would still be compatible with $H$ furnishing $0.1\% \lesssim \Omega h^{2} < 1\%$ (see also Ref.~\cite{Cline:2013bln}). Finally, we note that very large mass splittings, approaching the upper boundary of the shown parameter space, correspond to low transition temperatures $T_c \sim T_{\rm f.o.}$ and should be taken with a grain of salt, in particular because DM freeze-out could occur before the EWPhT. In this case, the longitudinal $Z\left( W \right)$ components are replaced by the scalar fields $\phi^{(\pm)}$ and it may be worth investigating the effect on DM annihilations. We leave a dedicated analysis for future work.

\section{Conclusion}

\label{sec:concl}

In this article, we investigated the prospects to realize the found DM abundance in the IDM, while taking theoretical and the latest experimental constraints into account, considering also regions of parameter space not explored in detail before. We found that the tight limits from XENON1T constrain the DM mass to the range $55\GeV \lesssim m_{H} \lesssim 75\GeV$ together with a small Higgs portal coupling $\left\vert \lambda_{345} \right\vert \lesssim 0.01$ or to a high mass regime of $m_H\gtrsim 500\GeV$. In the low mass range, we found a new viable spectrum for a realistic DM abundance, considering a small, but non-universal, splitting of $\Delta m = (8,25)\GeV$.
To arrive at this conclusion, the behavior of the cross sections of several (co-)annihilation channels was analyzed in detail. 

Afterwards, the surviving parameter space was studied in light of the EWPhT and we chose four BMs for an in-depth analysis of the evolution of the vacuum while the temperature decreases. The dependence of the EWPhT strengths $\xi_{\left( j \right)}$, both on the mass splitting and the coupling parameter $\lambda_{2}$, along with the behaviour of the transition temperature for different mass splittings were examined in detail. For the light-DM scenario, we discovered a broad mass range for the non-DM $\mathbb{Z}_{2}$-odd particles that leads to a strong first-order EWPhT either via one step or via two steps during the evolution of the universe. For the heavy-DM regime the phase transition seems generically to be second-order, unless $H$ only represents a sub-population of DM.
These quantitative findings, and in particular the characteristic two-step EWPhT, can be probed by investigating GW signatures. A detailed analysis in this direction is left for future work.

\acknowledgments
We are grateful to Andrei Angelescu, \'Alvaro Lozano Onrubia, Mar\'ia Dias, and Susanne Westhoff for useful discussions and comments. S. F. is grateful to the Heidelberg University and to the Heidelberg Graduate School for Physics. Y. J. is supported by the GuangDong Basic and Applied Basic Research Foundation (No. 2020A1515110150).

\bibliographystyle{JHEP}
\bibliography{RefsIDM}

\end{document}